\documentclass[aip,jcp,a4paper,reprint]{revtex4-1}
\usepackage{graphicx,graphics}
\usepackage{amssymb,amsmath,amsfonts}
\usepackage{array}

\begin{document}

\title{Model for Triplet State Engineering in Organic Light Emitting Diodes}

\date{\today}

\author{Suryoday Prodhan}
\email[Electronic mail: ]{suryodayp@gmail.com}
\affiliation{Solid State and Structural Chemistry Unit, Indian Institute of Science,
Bangalore 560012, India}
\author{Zolt\'{a}n G. Soos}
\email[Electronic mail: ]{soos@princeton.edu}
\affiliation{Department of Chemistry, Princeton University, Princeton,
New Jersey 08544, USA}
\author{S. Ramasesha}
\email[Electronic mail: ]{ramasesh@sscu.iisc.ernet.in}
\affiliation{Solid State and Structural Chemistry Unit, Indian Institute of Science,
Bangalore 560012, India}

\begin{abstract}
Engineering the position of the lowest triplet state $(T_1)$ relative to the first
excited singlet state $(S_1)$ is of great importance in improving the efficiencies
of organic light emitting diodes and organic photovoltaic cells. We have carried out
model exact calculations of substituted polyene chains to understand the factors that
affect the energy gap between $S_1$ and $T_1$. The factors studied are backbone
dimerisation, different donor-acceptor substitutions and twisted geometry. The largest
system studied is an eighteen carbon polyene which spans a Hilbert space of about
$991$ million. We show that for reverse intersystem crossing (RISC) process, the best
system involves substituting all carbon sites on one half of the polyene with donors
and the other half with acceptors.
\end{abstract}

\maketitle

\section{Introduction}

Conjugated polymers have become one of the prominent candidates in  
flexible, solid state organic light emitting diode (OLED) devices 
\cite{burroughes}. They are now employed in commercial displays and
lighting applications. Emission properties in these devices are 
primarily based on injection of an electron and hole from electrodes
into the device. These charges migrate under the influence of the
electric field and could finally recombine, giving rise to a singlet
or a triplet exciton on the conjugated system. Although simple spin
statistics predicts generation of at most $25\%$ singlet excitons
due to independent injection of electron and hole, this is not borne
out experimentally. The spin statistics does not account for the
rate of formation of the excitons, and the rate depends upon the
binding energies of excitons. The singlet exciton binding energy
being smaller than that of the triplet exciton, more singlets are
formed in a unit time than triplets. The exciton binding is purely a
consequence of electron-electron interactions and the $25\%$ upper
bound for singlet exciton formation is valid only in the 
noninteracting picture \cite{wohlge,tandon}. Notwithstanding this,
the actual yield of singlets is still small and low internal
efficiency of electroluminescence in OLEDs beckons alternate routes
such as harnessing triplet excited states. Inclusion of heavy metal
atoms like platinum (Pt) or iridium (Ir) in the conjugated polymer
enhances spin-orbit coupling which breaks the spin symmetry, thus
allowing what is notionally a singlet-triplet transition, indeed
electrophosphorescent devices have been developed 
\cite{balnat98,obrienapl99,balapl99} using this principle. However,
the longer lifetime of phosphorescence results in saturation of
triplet state population of the emitter and promotes triplet-triplet
annihilation (TTA). TTA could give rise to a lower energy nonemissive
state which will not contribute to light emission \cite{obrienapl99}.
Hence, the alternate pathway i.e. population enhancement via
conversion of triplet states into singlet states seems to be a
promising option for harnessing triplets to enhance
electroluminescence quantum yield. The position of triplet $T_1$
relative to the singlet $S_1$ is also of importance in other
applications such as photodynamic therapy, where collision between 
triplet oxygen and $S_1$ will give rise to $T_1$ and singlet oxygen,
the latter being the reactive species in the therapy.

In organic systems $T_1 \rightarrow S_1$ population transfer can be
achieved either through triplet-triplet annihilation or through
reverse intersystem crossing (RISC). Triplet-triplet annihilation
can theoretically maximize yield up to $\sim 11\%$ based on the
simplistic argument that two spin-1 species can give rise to one
spin two, one spin one and one spin zero species, resulting in a
theoretical maximum yield of $1/9$. This picture will change in
interacting models depending upon the exciton binding energy of the
species. However, there are claims that TTA can enhance OLED
efficiency up to $\sim 62.5\%$ \cite{gounatphot12,kondajap09}.
The excited singlet population can also be enhanced by the RISC 
mechanism. Although $E(T_1)<E(S_1)$ by Kasha rule, normally 
$E(T_1)$ is far less than $E(S_1)$ and triplets are lost. The 
RISC idea is to find systems in which the $S_1-T_1$ gap is of order
$k_BT$ under ordinary conditions. Thermal equilibrium may then
repopulate $S_1$ and depending upon competing processes, make
fluorescence the dominant decay mode for $T_1$ as well. Even better
would be violation of Kasha rule, molecules with $E(T_1)>E(S_1)$.
Thus reduction of energy gap between singlet excited state and
triplet excited state and utilization of environmental thermal
energy for RISC appears to be an attractive alternate path for
enhancing the efficiency of electroluminescent devices.

Experimental and theoretical investigations of RISC are being
carried out through the last decade, both in metal-containing and
all-organic (metal-free) conjugated molecules and oligomers. Kohler
et al. studied $S_1-T_1$ gap using fluorescence and
delayed-fluorescence techniques in platinum containing phenylene
ethylene polymers with spacers of different size and in their
all-organic analogs; corresponding energy gaps are of the order
$0.7\pm0.1$ eV irrespective of the organic ligand used
\cite{kohlerjcp02}. In spite of this success, interest in metal
containing polymers has waned since the metals in these systems are
usually rare-earth. Metal-free thiophenylene based copolymers
with (i) para-phenylene, (ii) ethylene, (iii) phenylene vinylene
and (iv) thioenylene vinylene moieties in their structure are
synthesized by Chaudhuri et al. \cite{chauangew10}. The lowest
achievable $S_1-T_1$ gap reported is as small as 0.02 eV, as
determined by the difference in the peak position of the
fluorescence and phosphorescence spectra \cite{chauangew10}. 
Endo et al. developed molecular luminophores where minimal spatial
overlap of frontier molecular orbitals (HOMO and LUMO), residing
on donor and acceptor moieties results in a gap of 0.11 eV
\cite{endoapl11}, while Uoyama and coworkers \cite{uoyanat12} have
reported better molecular system with lower than $100$ meV gap. In
both reports, it is proposed that, introduction of steric hindrance
results in very low spatial overlap between HOMO and LUMO of the
corresponding molecules and consequently reduced gap between the
excited states. This assumes that the energy difference between
$S_1$  and $T_1$ is governed by the exchange integral involving the
HOMO and LUMO orbitals, which to a first approximation is governed
by the differential overlap of the HOMO and LUMO orbitals. Goushi
et al. developed electroluminescent devices based on exciplex
formation between donor and acceptor molecules, corresponding
energy separation between $S_1$ and $T_1$ in these systems being
$\sim50$ meV \cite{gounatphot12}. Recently, Adachi et al.
synthesized molecules belonging to carbazole-triazine family with
smaller $S_1-T_1$ gaps ($~0.04$ eV) \cite{adaapl12} and they have
come up with some systems having comaparable gaps \cite{adajacs12},
in the amine-sulphone family.

Theoretical modelling of excited singlet-triplet gap in molecular
systems having a donor and acceptor moieties is done in configuration
interaction picture with only single particle-hole excitations.
Gierschner et al. investigated carbazole-paraterphenyl systems with
different substituents on donor and acceptor parts and with
different linkers between parent moities using time-dependent density functional
theory (TD-DFT) technique \cite{gierorel12}. However, TD-DFT still suffers
from lack of accurate functionals for the calculations. Kohler et
al. studied $S_1-T_1$ gap in planar and twisted conformations of
long $\pi-$conjugated oligomers of poly(p-phenylene vinylene) (PPV),
poly(p-phenylene) (PPP) and poly(p-phenylene ethynylene) (PPE)
\cite{kohleradfuncmat04}. These calculations were performed in
single CI space within the intermediate neglect of differential
overlap (INDO) model. They found that the energy gap between
singlet excited state $(S_1)$ and triplet state $(T_1)$ is
independent of structure and consistent with the value of $0.7$ eV,
as they argued that the exchange interaction is short ranged in
character, depends only on the electron-hole wavefunction overlap
and therefore will be invariant in longer chains. The twist of
each monomer about its neighboring monomer also does not affect
much as twist in general remains in the range of $\leq 40^{\circ}$.
Their calculation are based on the crude single CI approximation
whose validity is in question in the twisted conformation. There is
also work of Karsten et al. who have reported oligomers of
5,7-bis(thiophen-2-yl)thieno[3,4-b]pyrazine where the $S_1-T_1$ gap
reduces from $\sim 0.9$ eV to $\sim 0.5$ eV in the pentamers. The
theoretical calculation at the INDO level, predicts the gap to be
of the order of $\sim 0.8$ eV in the pentamers \cite{karstenjpca08}.
Although, a number of materials with low singlet-triplet gaps are
studied, most of them are molecular or oligomeric systems. Organic
electronic devices use both small molecules and large oligomers. 
Both have their own advantages and disadvantages. Processing of large
oligomers is an advantage but variation in oligomer structure from batch
to batch is a disadvantage. On the other hand, molecules have well-defined 
structure but their processing is not as simple as those of oligomers
and they also tend to crystallize, degrading the device performance.
In the case of small molecules intermolecular charge separation or 
exciplex formation is necessary for obtaining smaller gap systems and
this is possible only on introducing very strong donors and acceptors.

Our goal in the present paper is to explore various factors such
as the strength of electron correlations, role of donor and
acceptor substitutions, length of $\pi$-conjugation and the
geometry of the conjugated back-bone on the $S_1-T_1$ gap in
simple substituted and unsubstituted polyenes. Attributing the
$S_1-T_1$ gap to the strength of the exchange integral involving
HOMO and LUMO orbitals is equivalent to single CI approximation,
which is grossly inadequate in strongly correlated systems such as
the conjugated $\pi$-systems. Instead, in our approach we carry out
full CI calculation on model polyenes to explore the relative
importance of different factors that control the $S_1-T_1$ gap.
While unsubstituted polyenes in the polymer limit do not 
fluoresce strongly, substituted polyenes in principle show strong 
fluorescence, hence we have used substituted polyenes as model compounds 
in the study.
We have employed the Pariser-Parr-Pople model to study the
$S_1-T_1$ gap. We have also employed the Hubbard model study to
verify the accuracy of our extrapolations. In the PPP 
model, the ground state $S_0$ of an unsubstituted polyene consists
of predominantly singly occupied $p_z$ orbitals, while the singlet
$S_1$ state, if it is dipole allowed, has more contribution from a
pair of doubly occupied and empty $p_z$ orbitals. If $S_1$ is a
two-photon state, as is the case of long polyenes, it has more
probability for singly occupied states than even the ground state.
The triplet $T_1$ state is much like the $S_0$
state, except that the electron delocalisation is reduced due to
Pauli blocking for electron transfer between neighboring orbitals
having the same spin. In this study, we explore ways of increasing
ionicity of the $T_1$ state to raise its energy close to that of
the $S_1$ state, when $S_1$ is dipole allowed.

This paper is organized as follows. In Sec. II we introduce the model 
Hamiltonian and the methodology of our study. In Sec. III, we discuss 
the role of (i) conjugation length, (ii) dimerisation strength, 
(iii) donor-acceptor strength in push-pull systems, and (iv) the role 
of molecular geometry on the S1 − T1 gap, respectively. In Sec. IV, we 
summarize and conclude our study.

\section{Model Hamiltonian and Computational Method}

\begin{figure}[t]
\includegraphics[width=8.0cm]{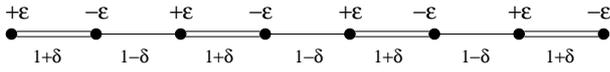}
\caption{\label{system} Schematic diagram of a polyene chain; $+\epsilon$ and
$-\epsilon$ represent the donor and acceptor substitutions
respectively. $\delta$ is the dimerisation parameter and the transfer
integral is modulated as $(1+\delta)t_0$ for the double bond and 
$(1-\delta)t_0$ for the single bond.}
\end{figure}

In the present model system calculation, we consider linear even polyene
chains of length varying between $4-18$ sites in steps of $2$ carbon
sites, so as to have even number of electrons in the $\pi$ system. In
substituted polyenes, three type of substitutions are
considered; in one, the donor and acceptor effects are introduced
alternately along all carbon atoms in the chain, while in the 
second we have considered the donor and acceptor substitutions
at the terminal carbon atoms and have rotated the molecule about the
middle bond, in $4n+2$ (n integer) polyene chain. In the third,
we have substituted one half of the $4n$ site polyene chain by 
donors and the other half by acceptors. The basic idea, behind
considering alternate donor-acceptor substitutions is to reduce the
effect of strong electron-electron interaction, for it is known that
in non-correlated picture the lowest singlet excited state $S_1$ and
lowest triplet excited state $T_1$ are degenerate. In Fig.
\ref{system}, we have schematically shown a polyene system with
donor-acceptor substitution at alternate sites. A positive site
energy corresponds to a donor group and negative for the acceptor 
group. In the absence of any substitution, all the carbon atoms are
taken to be identical, with site energy zero providing the reference
scale for strength of substitution.

The Hamiltonian employed for interacting $\pi$-electronic system is
the Pariser-Parr-Pople(PPP) Hamiltonian \cite{pariser-parr,pople},
which considers long-range Coulombic interaction along with on-site 
Hubbard interaction $(U)$:

\begin{equation}
\begin{split}
H_{PPP} &=\sum_{i,\sigma} t_0(1-(-1)^i \delta)(\hat c_{i,\sigma}^\dagger \hat c_{i+1,\sigma} +\mbox{ H.C.})+
\sum_i \epsilon_i \hat n_i \\ &+\sum_i \frac {U_i}{2} \hat n_i (\hat n_i-1) + \sum_{i>j} V_{ij}(\hat n_i-z_i)(\hat n_j-z_j) 
\end{split}
\label{ppp}
\end{equation}

\noindent
$\epsilon_i$ is the site energy at site $i$, $t_0$ is the mean
nearest-neighbor hopping integral, $U$ is the on-site Coulomb
interaction energy. The intersite interaction energies, $V_{ij}$
are obtained from Ohno interpolation scheme \cite{ohno}, assuming a
mean C-C bond distance of $1.4$\AA. The quantity $\delta$ is the
fraction of dimerisation which we varied between $0$ to $0.25$, the
C-C distance accordingly varies as $1.4(1 \pm \delta)$\AA, with 
$\delta$. The $\hat c_{i,\sigma}^\dagger$ ($\hat c_{i,\sigma}$)
operators create (annihilate) an electron with spin $\sigma$ in the
$p_z$ orbital at the $i^{th}$ carbon atom; $\hat n_i$ is the
corresponding number operator and $z_i$ is the local chemical
potential given by the number of electrons in orbital `i' that leave
the $i^{th}$ site neutral; for carbon in $\pi$-conjugation $z=1$.
The standard PPP Hamiltonian parameters
for carbon, namely $t_0=2.4$ eV and $U=11.26$ eV are chosen for our
study. Neglecting the last term, which is the intersite interaction
term leads to the Hubbard model, the parameter $U/t$ is usually
varied to model different interaction strengths in this model.

The PPP Hamiltonian being non-relativistic, conserves total
spin. Since we are interested in the singlet and triplet states, it
is best to solve for the eigenvalues in a spin adapted basis. This
has the twin advantage of dealing with smaller Hilbert space as well
as labelling the eigenstates by the total spin. We have employed the
diagrammatic valence bond (DVB) basis as the spin adapted basis and
the Rumer-Pauling rule to weed out linear dependence \cite{ramasesh}.
The resulting basis is linearly independent but non-orthogonal.
The VB basis can be easily generated and manipulated by using a 
bit representation scheme. This allows for handling very large basis sets 
encountered in full CI calculations of systems with up to $18$ orbitals at
half-filling \cite{ramasesh}.
The Hamiltonian matrix in this representation is non-symmetric. We use
the Rettrup's modification of the Davidson's algorithm for obtaining
a few low-lying states \cite{david,ret}. Since we express the
Hamiltonian matrix in a complete basis, the results obtained are
exact or the full CI results. The major drawback with this method is
that the full CI space becomes exponentially large with increase in
system size. The largest space we have worked with in this paper is
the triplet state of a polyene with $18$ carbon atoms which spans a
space of dimension $901,995,588$. The use of electron-hole (e-h)
symmetry and $C_2$ symmetry leads to subspaces of dimensionality about
one-fourth of this. However, both symmetries are killed if we
introduce non-zero site energies to simulate donor (acceptor)
behavior at the sites. Nonetheless, if we introduce non-zero site
energies as in Fig. \ref{system}, $C_2 \otimes e-h$ symmetry is
retained and exploiting this results in dimensionality of the 
subspaces which are half the dimensionality of the unsymmetrized space.
It should be remarked here that although the size of the
resultant matrices are large, the matrices are extremely sparse and
with Rettrup's algorithm, we can obtain a few low lying states in
each of the subspaces.

\begin{figure}[t]
\includegraphics[width=7.5cm]{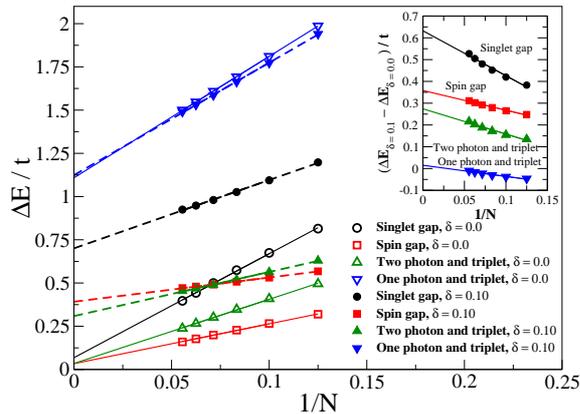}
\caption{\label{plot1bynhubb} Variation of different energy gaps in a regular Hubbard
chain ($\delta=0.0$; open symbols) and in a dimerised Hubbard chain
($\delta=0.10$; solid symbols) with number of sites $(N)$. The
energy gaps are represented according to the following: Singlet gap
$[E(S_1)-E(S_0)]$ (black circles); Spin gap $[E(T_1)-E(S_0)]$
(red squares); the gap between triplet state and two photon singlet
state $[E(2A_g)-E(T_1)]$ (green up triangles); gap between triplet
and one photon singlet state $[E(1B_u)-E(T_1)]$(blue down
triangles). Inset: Difference in the gap between $\delta=0$ and
$\delta=0.10$ is shown.}
\end{figure}

\section {Results and Discussion}

Spin is conserved in all the model systems that we discuss, but
other symmetries vary from model to model. The spin gap is always
the singlet-triplet gap, $E_{ST}=E(T_1)-E(S_0)$. We consider two
singlet-singlet gaps $E_1=E(S_1)-E(S_0)$ and $E_2=E(S_2)-E(S_0)$
and the crucial $S_1-T_1$ gap, $E(S_1)-E(T_1)$, which becomes
negative when Kasha rule is violated. 

\subsection{Unsubstituted uniform $(\delta=0)$ Hubbard and PPP models}

The Hubbard model in the $U=0$ limit is the noninteracting H\"uckel
model in which the $S_1-T_1$ gap is zero.
\vspace{1.5mm}

\begin{figure}[h]
\includegraphics[width=7.0cm]{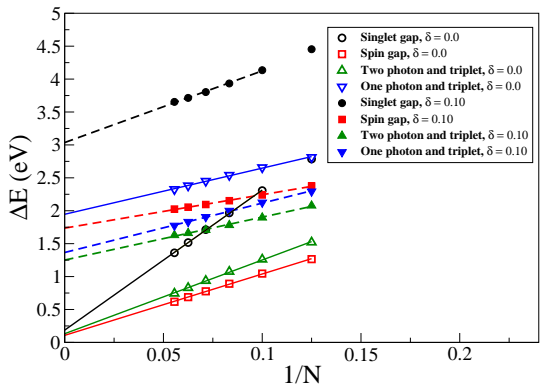}
\caption{\label{plot1bynppp} Variation of different energy gaps in a regular PPP chain
($\delta=0.0$, open symbols) and in a dimerised PPP chain
$\delta=0.07$, solid symbol) with site
Number $(N)$. The C-C bond lengths used for the dimerised chain
are $1.35$\AA and $1.45$\AA. The symbol code is given in the inset.
The definition of the gaps are the same as for
Fig. {\ref{plot1bynhubb}}.}
\end{figure}

\noindent
In the polymer limit of the
the uniform H\"uckel model, the gap from the ground state to the
first excited state is also zero since the system will be a half
filled one-dimensional band. When the Hubbard interaction $U$ is
turned on, then in the opposite limit, namely the
$U/t \rightarrow \infty$, the $S_1-T_1$ gap as well as the spin gap
$(S_0-T_1)$ vanish. The reason being, for the uniform Heisenberg
chain, the spin gap as well as the gap to the first excited singlet
state are zero. This also holds for the uniform Hubbard model in
the polymer limit.
In Fig. \ref{plot1bynhubb}, we see that the gap between the $2A_g$
two photon state and the lowest triplet are vanishingly small, with
the small extrapolated value indicating the magnitude of error in
extrapolation from finite systems to the polymer limit. However,
the gap between the one-photon state and $T_1$ remains finite. Our
Hubbard model calculation are carried out at $U/t=4$ and the two
photon state is below the one photon state. For this interaction
strength, the Hubbard chain will not be fluorescent, by Kasha rule.
The Hubbard model extrapolations of excited state energies give
results consistent with the physical picture \cite{liebwuprl68}.

\begin{figure}[b]
\includegraphics[width=7.5cm]{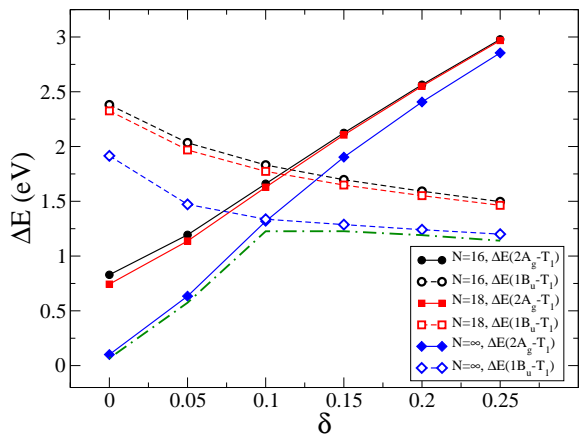}
\caption{\label{tdlimitvd} Variation of $2A_g-T_1$ gap (filled symbols) and
$1B_u-T_1$ gap (open symbols) for unsubstituted PPP chains with
dimerisation strength $(\delta)$. The symbol code is given in
the inset. The green curve shows the $S_1-T_1$ gap in the polymer
limit, independent of the symmetry of the $S_1$ state.}
\end{figure}

In Fig. \ref{plot1bynppp}, we show the dependence of $S_1-T_1$ gap
on chain length for uniform $(\delta=0)$ polyenes in the PPP model.
We see that the $S_1-T_1$ gap remains finite in the polymer limit
and reflects the fact that in the PPP model, as the chain length
increases, the Hamiltonian incorporates interactions of longer range.
Besides the $S_1$ state in the PPP model is the two photon state 
and the RISC process can only populate the nonemissive state, even
if this gap is small. Therefore, we see that in unsubstituted
correlated models, the vanishing of $S_1-T_1$ gap will not result
in a RISC process that can be useful in light emission.

\subsection{Dependence of $S_1-T_1$ gap on strength of dimerisation,
$\delta$}

\begin{figure}[t]
\includegraphics[width=8.0cm]{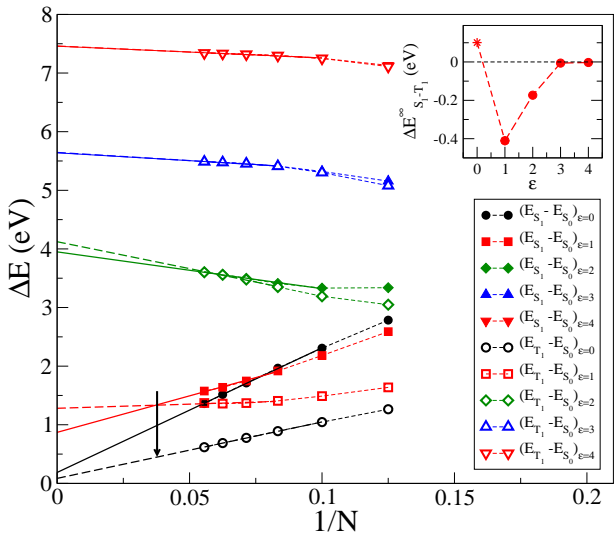}
\caption{\label{plot1bynd0} Variation of singlet gap [$E(S_1)-E(S_0)$; filled symbols]
and triplet gap [$E(T_1)-E(S_0)$; open symbols] with inverse chain
length $N^{-1}$ in unsubstituted and substituted regular PPP chain.
The symbol code for strength of the substitution is shown in the
figure. The chain length at which singlet gap is less than the
triplet gap is marked by arrow for $\epsilon=1$. Inset: $S_1-T_1$ gap
vs $\epsilon$ in the polymer limit.}
\end{figure}

In the noninteracting limit, even for nonzero $\delta$, the $S_1-T_1$
gap will be zero, although the $S_0-S_1$ and $S_0-T_1$ gaps remain
finite. At intermediate correlation strengths in the Hubbard model,
the $S_1-T_1$ gap corresponds to the gap between the one photon
state and the triplet state, since the gap will be dominated by the
transfer energy contribution to the excited state. However, at
$U=4t$, where we have studied, the lowest singlet is the two photon
state. The gap between the two photon singlet and triplet state
increases for $\delta=0.1$, compared to the uniform chain at every
system size and in the polymer limit gives a finite gap of
$\sim 0.25t$ (Fig. \ref{plot1bynhubb}). Thus the $S_1-T_1$ gap for
the dimerised chain is finite, unlike with the uniform Hubbard model.

In the PPP model also the $S_1-T_1$ gap increases with dimerisation
$\delta$ (Fig. \ref{tdlimitvd}). We see from the figure that both in
the large oligomers and in the polymer limit, there is a crossover
in the $2A_g$ and $1B_u$ states for $\delta \gtrsim 0.1$, as we
noted in {\cite{soosprl93}}. The gap between the lowest excited
singlet and the $T_1$ state is shown in the polymer limit
(bottom curve). The gap between $S_1$ and $T_1$ states is nearly
constant for $0.1<\delta <0.25$. These results show that
dimerisation alone is not a useful parameter for engineering the
$S_1-T_1$ gap.

\subsection{Dependence of $S_1-T_1$ gap on substitution}

We have studied the dependence of the $S_1-T_1$ gap on the strength
of substitution. We have simulated alternate substitution of donor
and acceptor groups of equal strength by introducing site energies;
positive site energy $(+\epsilon)$ at donor site and negative site
energy $(-\epsilon)$ at acceptor sites, $\epsilon >0$, both of same strength
$(\epsilon_D=-\epsilon_A)$. We have assumed four different donor
(acceptor) strengths by varying $\epsilon$ from $1$ eV to $4$ eV.
For the uniform chain we find that in the polymer limit (Fig.
\ref{plot1bynd0}), the $S_1-T_1$ gap nearly vanishes for
$\epsilon=3$ eV and $4$ eV. The $S_1-T_1$ gap is positive
$(E(S_1)>E(T_1))$ for $\epsilon=0$ (unsubstituted) case but for
$\epsilon=1$ eV, the triplet $T_1$ energy is higher than the
singlet $S_1$ energy, in the polymer limit, although
$E(N,S_1)>E(N,T_1)$ for $N$ value ranging from $8$ to $18$ that
we have studied. Indeed, the fact that for oligomers of
length $\gtrsim 30$ sites (see Fig. {\ref{plot1bynd0}}), the Kasha rule is not obeyed is an
advantage as $T_1$ to $S_1$ conversion will not need any thermal
energy. Thus such oligomers would be ideally suited as high
efficiency OLED materials. These results show that donor-acceptor
substitution is a very sensitive way to control the $S_1-T_1$ gap.

\begin{figure}[t]
\includegraphics[width=7.5cm]{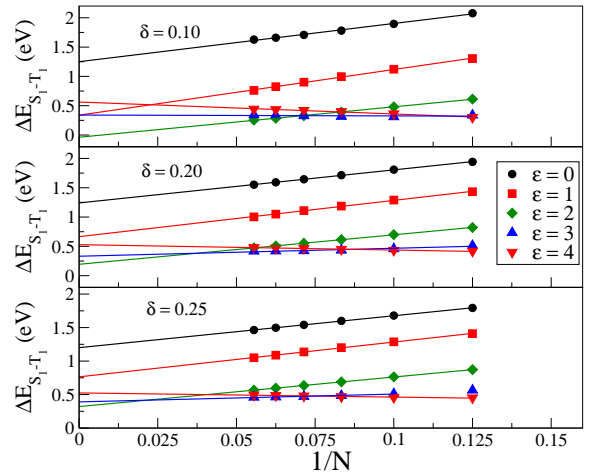}
\caption{\label{plot1bynd} Variation of $S_1-T_1$ gap with chain length $(N)$ in
unsubstituted and substituted dimerised PPP chains with dimerisation
constant $\delta=0.10, 0.20$ and $0.25$. Symbol code is given in
the middle panel.}
\end{figure}

\begin{figure}[b]
\includegraphics[width=5.0cm]{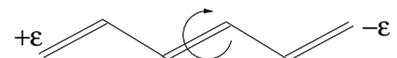}
\caption{\label{twsted} Schematic representation of $4n+2$ polyene with a
twist of angle $\theta$, around the middle bond. The terminal carbon atoms are substituted
by a donor and an acceptor. $|\epsilon|$ is measure of the donor (acceptor) strength and shifts
the orbital energy at the site by $\epsilon$ eV.}
\end{figure}

To explore the role of dimerisation in substituted chains, we have
studied the $S_1-T_1$ gap as a function of dimerisation of the PPP
chains with different substitution strengths. One of the important
features we note is that the $S_1-T_1$ gap in the polymer limit is
always positive, except in the case of $\delta=0.1$ and
$\epsilon=2$ ev, where the extrapolated gap is slightly below zero
(Fig. \ref{plot1bynd}). For all values of $\delta$, the $S_1-T_1$
gap extrapolates to the least value for $\epsilon=2$ eV. This shows
that just moderately strong donor-acceptor substitution is
sufficient to bring the $S_1$ and $T_1$ states close in energy.

\subsection{Dependence of $S_1-T_1$ gap on structure}

\begin{figure}[t]
\includegraphics[width=8.0cm]{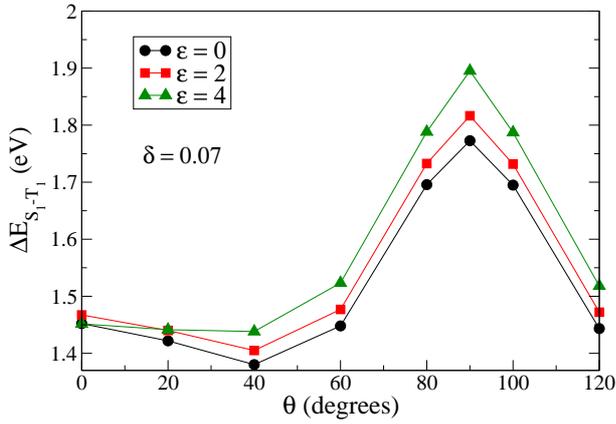}
\caption{\label{s1t1vd} Variation of $S_1-T_1$ gap with twist angle around the
middle bond $(\theta)$ in a linear polyene chain of $14$ sites in the
PPP model, unsubstituted or substituted at the ends by a donor and an
acceptor of equal strength $\epsilon$. The symbols are defined in the
inset and the dimerisation $\delta$ is taken to be $0.07$.}
\end{figure}

In correlated systems it is conjectured that the singlet excitation
is a charge like excitation which creates a pair of positive and
negative charges while triplet excitation involves creation of a
radical pair. This simple picture leads to the belief that if the
geometry of a polyene system is twisted, we can separate the charges
in the singlet exciton and spins in the triplet exciton, resulting
in a situation where these separated entities do not overlap. In
such a situation, it can be argued that the triplet state and the
excited singlet state should be very nearly degenerate. To test this
paradigm, we have studied the excited singlet and triplet states of
twisted polyenes with $4n+2$ carbon atoms with a donor and an
acceptor of equal strengths substituted at the end sites of the
chain, as a function of the twist angle (Fig. \ref{twsted}). 
The donor and acceptor are symmetrically placed so that we can exploit
the $C_2  \otimes e-h$ symmetry of the Hamiltonian. The twist is effected about
the central double bond and the transfer integral is taken as
$t\cos\theta$ where $\theta$ is the twist angle and $t$ is the
transfer integral which is $2.568$ eV, corresponding to a polyene
double bond. In Fig.\ref{s1t1vd} we have presented the $S_1-T_1$ gap
as a function of the twist angle for a polyene chain of $14$ carbon
sites, and for different site energies. We find that the $S_1-T_1$
gap is large and remains so as the central bond is twisted. The
dependence, though weak is nonmonotonic and shows a minimum around
$40^{\circ}$ twist angle. This result can be explained by the fact
that the charge separation leads to lower singlet excitation energy
but to vanishing triplet excitation energy. At $\theta=90^{\circ}$,
the gap between the ground state and the triplet excited state is
zero, but the gap between the ground state and the singlet excited
state though near a minimum is still very large since the
interaction between the charges at either ends $(U-V_{1N})$ is quite
large. Thus in strongly correlated systems, it is not possible to
reduce the $S_1-T_1$ gap by blocking the transfer between two halves
of the system, with the type of substitution that we have so far 
considered.  

\begin{figure}[t]
\includegraphics[width=8.0cm]{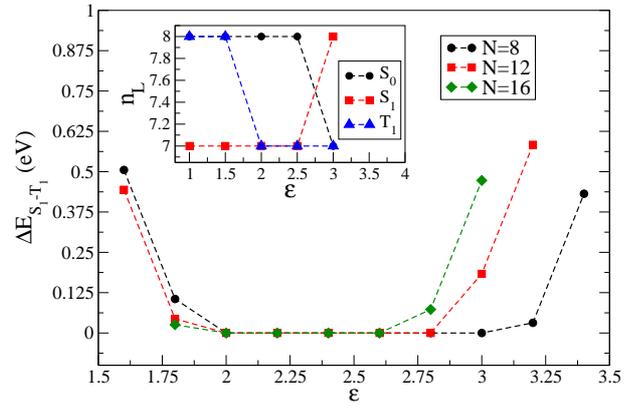}
\caption{\label{s1t1ve} Variation of $S_1-T_1$ gap with substitution energy 
$\epsilon$ in a linear dimerised polyene chain of $4n$ (n integer)
sites, one half substituted by donors and another half by acceptors
of equal strength, in the PPP model. Symbols indicating the chain
lengths are defined in the figure. The dimerisation factor $\delta$
is taken to be $0.07$ while the transfer energy $t$ for the central
bond is taken as zero. The inset shows the number of electrons, $n_L$
on the left half of the chain, for different states, as $\epsilon$
is varied $(N=16)$. The number of electrons on the right half, $n_R$ is 
$N-n_L$.}
\end{figure}

If the substitution on the polyene chain is such that both the excited
singlet and excited triplet states are ionic in the same way, we can in
principle reduce the $S_1-T_1$ gap. To test this, we have studied 
the $S_1-T_1$ gap in $4n$ (n integer) carbon
polyenes when on one half of the chain we have donors and on another
half we have acceptors (Fig. \ref{s1t1ve}). The chain is twisted about
the middle bond by
$90^{\circ}$ so that the transfer between the two halves of the chain
is zero. In this geometry, we have obtained the $S_1-T_1$ gap for
various donor/acceptor strengths, $\epsilon$. We find that the 
$S_1-T_1$ gap vanishes for a range $1.8\le|\epsilon|\le2.5$. 
In this range, the two halves of the chain are neutral in $S_0$ and degenerate 
ion-radicals in $S_1$ or $T_1$ since there is no hopping between the halves. 
$S_1$ is ionic and $T_1$ is neutral for $|\epsilon|<1.8$eV, while $T_1$ is ionic 
and $S_1$ is neutral for $|\epsilon|>2.5$eV. In either case, the $S_1-T_1$ degeneracy 
is lost.This can be seen in Fig. \ref{s1t1ve}
inset, where we have shown the number of $\pi$-electrons in the left half of
chain as a function of $\epsilon$, for $16$-site polyene chain. Indeed, some
of the experimental systems have the feature of donor substituted sites and
acceptor substituted sites connected through a twisted bond with 
very small transfer integral between the two substituted part \cite{chauangew10,endoapl11,uoyanat12}.

\section{Conclusion}

\begin{figure}[t]
\includegraphics[width=8.5cm]{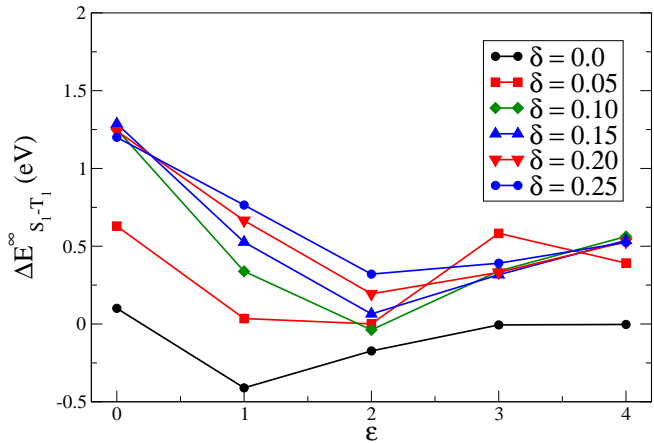}
\caption{\label{tdlimit} Dependence of the $S_1-T_1$ gap,
$\Delta E^{\infty}_{S_1-T_1}$, in the polymer limit, on $\epsilon$
and $\delta$, within the PPP model. The symbol code is given in the 
inset.}
\label{tdlimit}
\end{figure}

Engineering the energy gap between the triplet $(T_1)$ state and the
excited singlet $(S_1)$ state is of importance in improving the 
efficiency of organic electronic devices such as OLEDs and organic photovoltaic cells.
The aim of this paper has been to find the factors that affect the
gap between excited singlet $(S_1)$ and the lowest triplet $(T_1)$
state of a $\pi-$conjugated molecules. We have carried out exact or
Full CI calculations on polyene chains with up to $18$ carbon atoms.
We find that the usual factors such as change in dimerisation and
rotation about the central double bond do not materially affect
this gap. However, substitution by donor and acceptor groups
at alternate carbon sites has a strong effect on the $S_1-T_1$ gap
and the gap nearly vanishes for some values of the donor (acceptor)
strength and dimerisation parameter. Substitution with
donor/acceptor groups renders the triplet $T_1$ state more ionic in
character and therefore raises its energy closer to that of the
singlet $S_1$ state which is known to be ionic in character. This
study provides a basis for systematically controlling the
$S_1-T_1$ gap and will be useful in designing molecules with small
$S_1-T_1$ gap. Fig. \ref{tdlimit} summarizes the dependence of
$S_1-T_1$ gap on the factors such as dimerisation and substitution by
push-pull groups at alternate sites. However, the most promising case is
when we have donors substituted at all sites on one half of the chain and
acceptor substituted at the other half. In this case the $S_1-T_1$ gap
vanishes for a range of donor (acceptor) strengths, when the chain is twisted
around the middle bond separating the donors and acceptors. It should be
possible to synthesize such systems for device applications.
\vspace{0.5cm}

\begin{acknowledgments}
SR is thankful to the Department of Science and Technology, India
for financial support through various grants. SP acknowledges
CSIR India for a junior research fellowship.
\end{acknowledgments}

\end{document}